\documentclass[twocolumn]{jpsj2}

\usepackage{alltt}
\usepackage{graphics}

\newlength{\minipagetbllength}
\newcounter{minipagetblcounter}

\title{Effects of Velocity Correlation on Early Stage of Free Cooling Process of Inelastic Hard Sphere System}
\date{}

\makeatletter
\def\tr{\mathop{\operator@font tr}\nolimits}
\makeatother
\makeatletter
\def\Tr{\mathop{\operator@font Tr}\nolimits}
\makeatother

\def\runauthor{Ryo Kawahara and Hiizu Nakanishi}

\author{
 Ryo \textsc{Kawahara}\thanks{E-mail : ryokawa@stat.phys.kyushu-u.ac.jp} 
 and
 Hiizu \textsc{Nakanishi}\thanks{E-mail : naka4scp@mbox.nc.kyushu-u.ac.jp}}

\inst{Department of Physics, Kyushu University 33,
 Fukuoka 812-8581}

\recdate{\today}

\kword{
inelastic hard sphere system, free cooling, numerical simulation,
granular system
}

\abst{%
The free cooling process in the inelastic hard sphere system is
studied by analysing the data from large scale
 molecular dynamics simulations
on a three dimensional system.
The initial energy decay,
the velocity distribution function, and the velocity
correlation functions are calculated to be compared with theoretical
predictions.
The energy decay rate in the homogeneous cooling state is 
slightly but distinctively smaller than that expected 
from the independent collision assumption.
The form of the one particle velocity distribution is found not 
to be stationary.
These contradict to the predictions of the kinetic theory based
on the Enskog-Boltzmann equation and suggest that the 
velocity correlation is already important in the early stage of 
homogeneous cooling state.
The energy decay rate is analysed in terms of the velocity correlation.
}

\begin{document}

\pagestyle{plain}
\maketitle

\section{Introduction}
\suppressfloats[t]

The hard sphere system is the oldest system studied in the statistical mechanics and is still attracting much interest of current research. Some of the recent studies on this system were triggered by the recent appreciation of the fact that, no matter how small the inelasticity is, it could cause drastic changes in the system behavior and leads to series of transitions as the system evolves, as long as the system is large enough\cite{doc1:btwo_dim_regime}.

In the inelastic system, the particles lose relative velocity every time the particles collide each other. The first consequence of this is the uniform cooling of the system (Homogeneous Cooling State, HCS). In this stage, no macroscopic structure is assumed to exist and we can measure the kinetic energy in terms of the granular temperature $T$ defined by
\begin{equation}
\label{doc1:maverage_temperature}
T \equiv \frac{m}{d} \langle v^{2} \rangle,
\end{equation}
where $d$ is the system dimensionality, $m$ is the mass, and $\langle \cdots \rangle$ represents the statistical average. Then it has been shown that the initial cooling follows as
\begin{equation}
T(t) = \frac{T_{0}}{(1+t/t_{0})^2},
\end{equation}
where $t$ is the time, and $t_{0}$ and $T_{0}$ are constants. This is called Haff's law\cite{doc1:bhaff}.

Secondly, the velocity distribution deviates or does not converge to the Maxwell-Boltzmann distribution \cite{doc1:bvelocity_dist_hgf}. Theoretical analyses and numerical simulations have shown that the velocity distribution in an inelastic system falls into a distribution different from the Maxwell-Boltzmann within a first several collisions per particle if the system is uniform\cite{doc1:bgaussian}. This form of distribution has been shown to be a stable and stationary solution of the Enskog-Boltzmann equation of the kinetic theory for an inelastic system. Molecular dynamics simulations on a two dimensional system, however, have shown that this form of scaled distribution does not actually remain stationary but gradually returns back to the Maxwell-Boltzmann due to the velocity correlation ignored in the Enskog-Boltzmann approximation\cite{doc1:blaguerre2d}.

This velocity correlation in the inelastic system is developed by the reduction of relative velocity of colliding particles during the global cooling. Since the inelastic collision reduces kinetic energy but not momentum, random fluctuations in momentum over a certain spatial region come to stick out when the size of momentum fluctuations becomes comparable with the momentum itself. This is called the noise reduction instability\cite{doc1:bkinetic_theory}. The velocity correlation may be seen as a vortex structure in the velocity field of the system.

Well developed vortex structure causes shearing in flow between vortices, which in turn causes viscous heating. The pressure gets higher in the heated region, then this leads to particle flow due to the resulting pressure gradient, and eventually inhomogeneity of the system in the density field\cite{doc1:borigin_of_density_cluster}, which is called the clustering instability.

In this paper we present a detailed analysis of numerical simulations in a three dimensional system, and give evidences that velocity correlations have a deep influence on the behavior of inelastic hard sphere system even in a very early stage.

The paper is organized as follows; In \S\ref{doc1:s_model_system}, we introduce our model system and the simulation setup. In \S\ref{doc1:s_hcs}, we present our simulation result for the time dependence of the kinetic energy. In \S\ref{doc1:s_velocity_distribution}, the velocity distribution is compared to the one obtained from the theory based on the Enskog-Boltzmann equation using the generalized Laguerre polynomial expansion to show the deviation from the Gaussian. In \S\ref{doc1:s_vel_correlation}, we compare the velocity correlation with the theoretical result based on the hydrodynamic equation with the Langevin force. In \S\ref{doc1:s_energy_decay_rate}, the deviation of the energy decay rate from Haff's law is analysed in detail. We summerize our results in \S\ref{doc1:s_summery}.

\section{Model system}
\label{doc1:s_model_system}
\suppressfloats[t]

The model system we study is a three dimensional (3-d) hard sphere system which undergoes inelastic collisions with a constant normal restitution coefficient $e$. We consider the mono-disperse system with the diameter $\sigma$ and the mass $m$. The particle rotation is ignored for simplicity. Then the collision rule is given in terms of the velocity of the $i$-th particle $\mib{v}_{i}$ and $\mib{v}_{i}^{*}$ before and after the collision, respectively, as 
\begin{align*}
\mib{v}_{i}^{*} &= 
  \mib{v}_{i}
  - \frac{1+e}{2}[\mib{n}\cdot(\mib{v}_{i} - \mib{v}_{j})]\mib{n}, \\
\mib{v}_{j}^{*} &= 
  \mib{v}_{j}
  + \frac{1+e}{2}[\mib{n}\cdot(\mib{v}_{i} - \mib{v}_{j})]\mib{n},
\end{align*}
 where $i$ and $j$ are the colliding particles, and $\mib{n}$ is the unit vector parallel to the relative position of the colliding particles at the time of contact. In the following, we use the unit system where the particle diameter, the mass, and the initial average kinetic energy per particle are unity.

To simulate the system, we employ the event driven method, in which we keep track of colliding events. The actual code of the simulations is based on the fast algorithm by Isobe \cite{doc1:bsimple_efficient_algorithm}, which enables us to simulate the system with the particle number as many as $10^{6}$ in 3-d on our PC's. Most of the simulations presented here are done for the restitution coefficient in the range of $e \geq 0.8$ with the number density $n = 0.4$, which corresponds with the volume fraction $\nu\equiv(4/3)\pi (\sigma/2)^3 n = 0.209$. The density dependence is also simulated for some observables.

The system is contained in the cube with the periodic boundary condition. The initial state is prepared as the equilibrium state by running the system long enough with the restitution coefficient $e = 1$. To check the equilibration, we monitor the velocity distribution and wait until some of the Laguerre coefficients become zero within a statistical error; this procedure typically takes about 50 collisions per particle.

The inelastic collapse\cite{doc1:binelastic_collapse} does not occur without any special treatment in the present simulations because we mainly focus on the initial and intermediate stage of cooling with the parameters of relatively low dissipation.

\section{Energy decay in HCS}
\label{doc1:s_hcs}
\suppressfloats[t]

As the system cools, things slow down and colliding events become less and less frequent. Therefore, it is convenient to measure the time in terms of the average number of collisions $\tau$ that each particle experiences. This is related with the time $t$ by
\begin{equation}
d\tau = \omega dt ,
\end{equation}
where $\omega$ is the collision frequency, which depends on time.

In HCS, collisions are thought to be uncorrelated and the average energy loss in each collision is proportional to the kinetic energy, or the granular temperature, therefore we have
\begin{equation}
\label{doc1:m_temp_time}
\frac{dT}{dt} = -2\gamma\omega T ,
\end{equation}
where $\gamma$ is the energy loss rate. Within the approximation of random uncorrelated collision with the Maxwell-Boltzmann velocity distribution, $\gamma$ becomes
\begin{equation}
\label{doc1:m_gamma0}
2\gamma_{0} = \frac{1-e^{2}}{d},
\end{equation}
because only the normal component of the velocity reduces by the factor $e$. This gives Haff's law for Maxwell-Boltzmann velocity distribution,
\begin{equation}
\label{doc1:m_haff_tau}
T_{\mathrm{Haff}}(\tau) = T_{0} \exp[-2\gamma_{0}\tau].
\end{equation}

\begin{figure}[htb]
\begin{center}
\scalebox{0.45}{\resizebox{\textwidth}{!}{\includegraphics{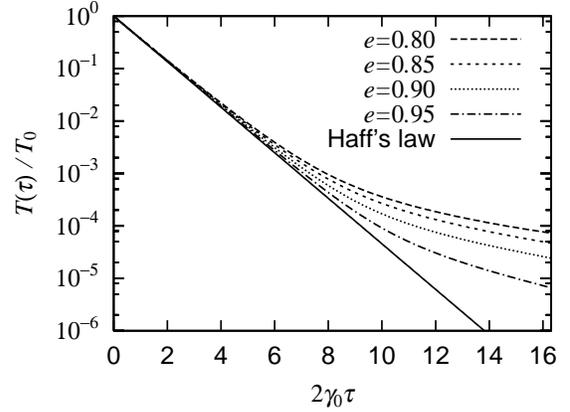}}}
\caption{The time dependence of the
granular temperature $T$ defined
in eq.(\ref{doc1:maverage_temperature})
for various restitution coefficients $e$
with $n = 0.40$
and $N = 6\times 10^{5}$.
The average number of collisions $\tau$
is scaled as $2\gamma_{0}\tau$ .
The solid straight line represents the
Haff's law with $\gamma_{0}$
(eq.(\ref{doc1:m_haff_tau})).}
\label{doc1:f_gamma_scaled_energy}

\end{center}
\end{figure}

Figure \ref{doc1:f_gamma_scaled_energy} shows the energy decay;
the granular temperature $T(\tau)$ defined
in eq.(\ref{doc1:maverage_temperature}), which is proportional
to the kinetic energy, is plotted 
as a function
of $2\gamma_{0}\tau$ for several values
of $e$.
Although the initial exponential
decay seems to be well represented by Haff's law
(\ref{doc1:m_haff_tau}),
there is a small but distinctive
discrepancy in the decay rate $\gamma$
from eq.(\ref{doc1:m_gamma0})
as we will see in Fig. \ref{doc1:f_gamma_int}
in \S\ref{doc1:s_energy_decay_rate},
where we plot numerically obtained decay rate $\gamma$ with the
theoretical analysis that we will discuss in detail.
There are two possible causes for this:
(i) the deviations of velocity distribution
    from the Maxwell-Boltzmann distribution,
(ii) the correlation in collisions. 
We will discuss (i) and (ii) in \S\ref{doc1:s_energy_decay_rate}.

\begin{figure}[htb]
\begin{center}
\scalebox{0.45}{\resizebox{\textwidth}{!}{\includegraphics{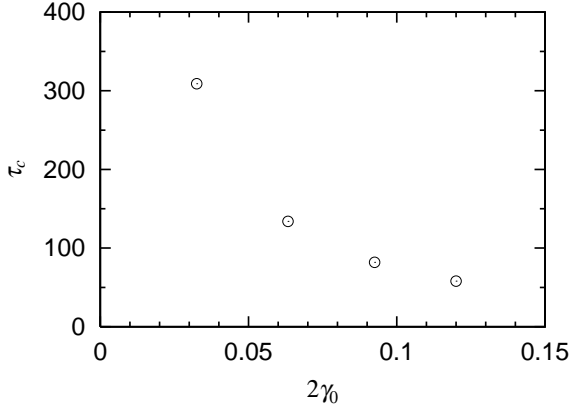}}}
\caption{The crossover time $\tau_{c}$ as a function
of $2\gamma_{0} = (1-e^{2})/d$.
The circles represent our simulation results
for $e = 0.95,\,  0.90,\,  0.85$,
and $0.80$ from left to right.}
\label{doc1:f_tau_c-3}

\end{center}
\end{figure}

The energy decay deviates from the exponential as the time elapses, after then the clustering occurs. We define the crossover time $\tau_{c}$ between Haff's region and the clustering region as the time when $T_{\mathrm{Haff}}(\tau)$ becomes half of $T(\tau)$, namely
\begin{equation}
\label{doc1:m_def_tauc}
\frac{1}{2}T(\tau_{c}) = T_{\mathrm{Haff}}(\tau_{c}) ,
\end{equation}
where $T_{\mathrm{Haff}}(\tau)$ is given by eq.(\ref{doc1:m_haff_tau}). In the case of $n = 0.40$, we obtain $\tau_{c} =  58,\, 82,\, 134,$ and $309$ for $e = 0.80,\, 0.85,\, 0.90,$ and $0.95$ respectively. The crossover time $\tau_{c}$ is plotted against $2\gamma_{0}$ in Fig. \ref{doc1:f_tau_c-3}. In the following sections, we will focus on the system behavior in early stage, $\tau \leq \tau_{c}$.

The beginning of the clustering has been observed around $\tau \sim \tau_{c}$ in our simulation and it has a strong effect on the energy decay. The long time behavior of the energy in the clustering region seems to be some power decay like $\tau^{-\alpha}$. From the hydrodynamical approximation, $\alpha = d / 2$ has been obtained\cite{doc1:bextension_haff}, but our result seems to show $\alpha = 3 \sim 4$ with some sample dependence and parameter dependence.

\section{Velocity distribution}
\label{doc1:s_velocity_distribution}
\suppressfloats[t]

The velocity distribution function $f(\mib{v}, \tau)$ in the inelastic system is conveniently represented in the scaled form using the average speed
\begin{equation}
v_{0}^{2}(\tau) = 
\frac{2}{d}\int d\mib{v} f(\mib{v},\tau)v^{2} ,
\end{equation}
as
\begin{equation}
f(\mib{v}, \tau) =
 \frac{1}{(v_{0}(\tau)\sqrt\pi)^{d}}\thinspace
 \rho(\mib{c}, \tau);
\quad
\mib{c} \equiv \frac{\mib{v}}{v_{0}(\tau)}.
\end{equation}
The scaled velocity distribution function $\rho(\mib{c},\tau)$ would be independent of $\tau$ if the form of the velocity distribution were stationary as the system cools down.

When the distribution function is not far from the Gaussian, it is reasonable to expand it as
\begin{equation}
\rho(\mib{c},\tau) =
 \exp(-c^{2})\sum_{\ell = 0}^\infty a_{\ell}(\tau)L_{\ell}^{\alpha}(c^{2}),
\quad (\alpha = \frac{d}{2} - 1),
\end{equation}
using the generalized Laguerre polynomials, or the Sonine polynomials,\cite{doc1:bvelocity_dist_hgf}\cite{doc1:bgaussian} $L_{\ell}^{\alpha}(x)$. The coefficients $a_{\ell}(\tau)$ can be obtained by
\begin{equation}
a_{\ell}(\tau) = 
 \frac{\ell !\thinspace \Gamma(d/2)}{{\sqrt\pi}^d\thinspace \Gamma(\ell+d/2)}
 \int d\mib{c}\rho(\mib{c},\tau) L_{\ell}^{\alpha}(c^{2}),
\end{equation}
where $\Gamma(x)$ is the gamma function. The values of the first two coefficients are given by
\begin{equation}
a_{0} = 1, \quad a_{1} = 0,
\end{equation}
due to the normalization and the scaling, but any deviation from the Maxwell-Boltzmann distribution is represented by non-zero values of the higher order coefficients $a_{\ell} \thinspace (\ell \ge 2)$.

The time development of $a_{\ell}(\tau)$ has been calculated based on the Enskog-Boltzmann equation with a certain truncation scheme for the higher order coefficients\cite{doc1:bgaussian}. It has been shown that the Enskog-Boltzmann equation has the solution which starts from $a_{0} = 1$ and $a_{i} = 0$ $(i \ge 1)$ and converges to non-zero constants within a several collisions per particle.

\begin{figure}[htb]
\begin{center}
\scalebox{0.45}{\resizebox{\textwidth}{!}{\includegraphics{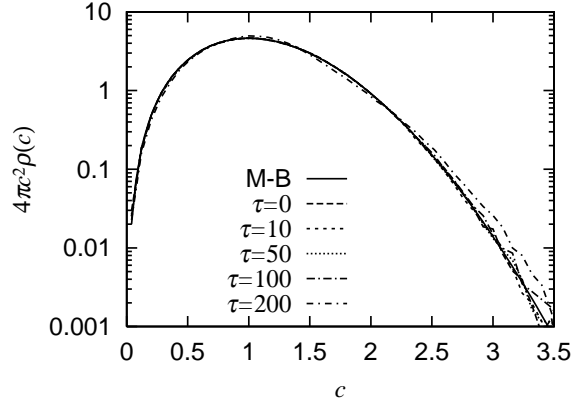}}}
\caption{The scaled velocity distribution
function $\rho(c)$
at a several values of $\tau$
for the system with
$e = 0.80,\, n = 0.40,\, N = 6\times 10^{5}\,$.
The crossover time $\tau_{c}$ is 58 for
this system.
The solid line represents the Maxwell-Boltzmann distribution
$\rho(c) = \exp(-c^{2})$.}
\label{doc1:f_veldist_scaled}

\end{center}
\end{figure}

\begin{figure}[htb]
\begin{center}
\scalebox{0.45}{\resizebox{\textwidth}{!}{\includegraphics{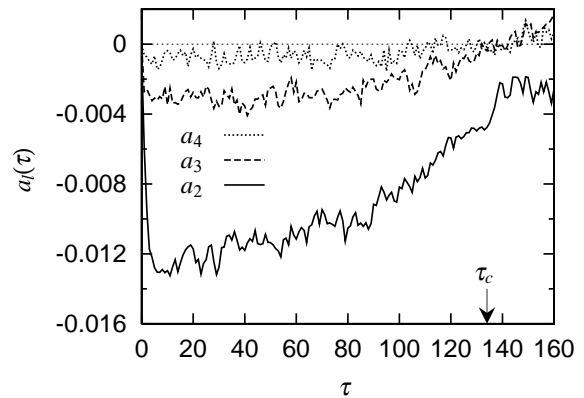}}}
\caption{The time development of the generalized Laguerre polynomial
expansion coefficient $a_{\ell}$
of the $\rho(c)$
for $e = 0.90,\, n = 0.40,\, N = 6\times 10^{5}$.
The data are averaged over three realizations.
The crossover time $\tau_{c}$ is
also shown by the arrow.}
\label{doc1:f_laguerre_al}

\end{center}
\end{figure}

\begin{figure}[htb]
\begin{center}
\scalebox{0.45}{\resizebox{\textwidth}{!}{\includegraphics{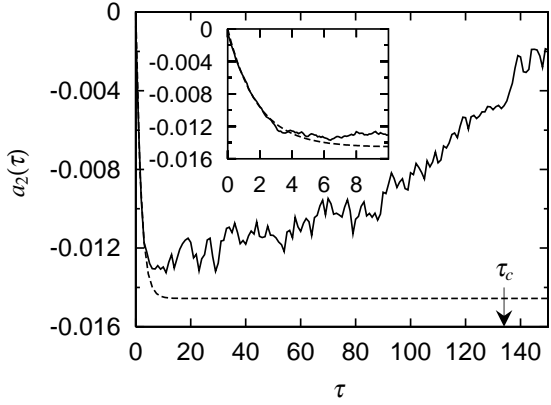}}}
\caption{Comparison of $a_{2}$ of the simulation
result (solid line) 
with the solution of the Enskog-Boltzmann equation with the second order
approximation  given in eq.(\ref{doc1:m_a2_analytical}) (dashed line)
for $e = 0.90,\, n = 0.40$
and $N = 6\times 10^{5}$
(averaged over 3 realizations).
The inset shows the short time behavior (averaged over 6 realizations).}
\label{doc1:f_laguerre_theory}

\end{center}
\end{figure}

Figure \ref{doc1:f_veldist_scaled} shows the scaled 
distribution of normalized particle
speed $c$ for
the system with $e=0.8$ for
a several values of the time $\tau$ in the
semilogarithmic scale.  The deviation from the Gaussian 
is very difficult to see even after the clustering sets in
around $\tau_{c} = 58$.

The Laguerre coefficients $a_{2}, \, a_{3}$, and $a_{4}$ are shown as functions of $\tau$ in Fig. \ref{doc1:f_laguerre_al} for $e = 0.90$ and $n = 0.40$. They are small but clearly have non zero values, which shows that the coefficients deviate from those for the Gaussian. The initial deviation from the Gaussian is very quick as has been predicted, and $|a_{2}| > |a_{3}| > |a_{4}|$ holds, which supports the validity of the truncation scheme. In Fig. \ref{doc1:f_laguerre_theory}, the simulation results are compared with the solution of the Enskog-Boltzmann equation for $a_{2}(\tau)$, whose expression is given in Appendix\ref{doc1:as_laguerre_coeff}. The simulation results agree very well with the solution till it becomes close to the stationary value, but after then $a_{2}(\tau)$ and other coefficients go back towards zero until $\tau\sim\tau_{c}$.

\begin{figure}[htb]
\begin{center}
\scalebox{0.45}{\resizebox{\textwidth}{!}{\includegraphics{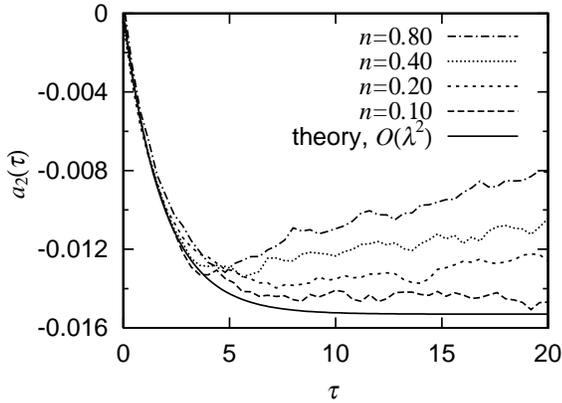}}}
\caption{The time dependence of $a_{2}$
for various values of the 
density $n$
with $e = 0.85$
and $ N = 6\times 10^{5}$.
The solid line represents the solution of the Enskog-Boltzmann equation with
second order approximation given in eq.(\ref{doc1:m_a2_analytical}).}
\label{doc1:f_laguerre_theory_short}

\end{center}
\end{figure}

\begin{figure}[htb]
\begin{center}
\scalebox{0.45}{\resizebox{\textwidth}{!}{\includegraphics{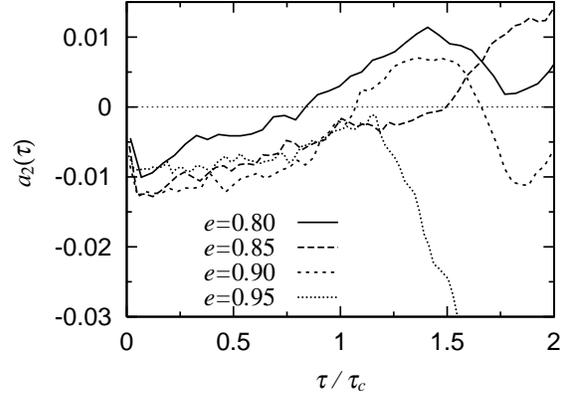}}}
\caption{Long time behavior of $a_{2}$ for
various values of the restitution coefficient $e$
at $n = 0.40$
and $ N = 6\times 10^{5}$.
The data presented in the graph are smoothed by the time average of one sample 
over the interval
$\Delta \tau = 3,\, 3,\, 5, \mbox{ and } 7 $
for $e = 0.80,\, 0.85,\, 0.90, \mbox{ and } 0.95 $,
respectively, from the original data being measured
at every integer $\tau$.}
\label{doc1:f_laguerre_e}

\end{center}
\end{figure}

The coefficients obtained from the simulations depend on the density although the Enskog-Boltzmann equation results do not. Figure \ref{doc1:f_laguerre_theory_short} shows the short time behavior of $a_{2}(\tau)$ for various densities $n$. $a_{2}(\tau)$ for the smaller density case is close to the theoretical solution. This suggests that this behavior of returning to the Gaussian is caused by velocity correlation because multiple collisions of the same pairs occur more frequently in higher density. Actually, it has been shown in the 2-d simulation that the Laguerre coefficients stay stationary if the velocity correlation is destroyed artificially by shuffling the velocity of each particle\cite{doc1:blaguerre2d}.

The behaviors of $a_{2}(\tau)$ for various restitution coefficients $e$ are shown in Fig. \ref{doc1:f_laguerre_e} with the time being scaled by the crossover time $\tau_{c}$. All of them show similar trend when the time is scaled with $\tau_{c}$. The behavior after the clustering ($\tau > \tau_{c}$) has large sample dependence. This seems to be due to the fact that the system is not large enough compared with the cluster size.

\section{Velocity field and its correlation}
\label{doc1:s_vel_correlation}
\suppressfloats[t]

The random velocity field in the initial state becomes organized during HCS through inelastic collisions. This has been clearly seen in 2-d system as the vortex structure \cite{doc1:bcahnhilliard}. In 3-d system, it is rather difficult to see directly due to a problem of visualization and also due to the difficulty to have a large system which contains enough vortices to see their structure.

\begin{figure}[htb]
\begin{center}
\scalebox{0.45}{\resizebox{\textwidth}{!}{\includegraphics{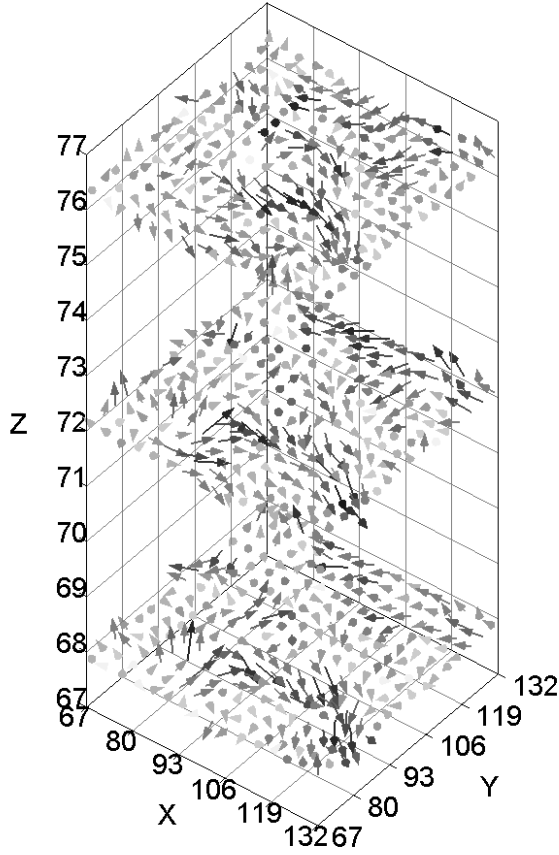}}}
\caption{Cross sections of velocity field
at $\tau = 60 \, (\tau_{c} = 58)$
for $n = 0.40,\,  e = 0.80$
and $N = 1\times 10^{6}$ with the system size $L = 136$.
Only $1 / 4$ of  the $xy$ plane
and $3 / 32$ of the $z$ direction of the
whole system are shown.
The presented velocity field is the average velocity over all particles
in a cell whose size is $1 / 32^{3}$ of the
system.
The tone and the length of the arrows represent the
magnitude of the velocity (the darker arrow represents larger velocity).}
\label{doc1:f_snapshot}

\end{center}
\end{figure}

Figure \ref{doc1:f_snapshot}shows a snapshot of a velocity field just before the clustering starts. The velocity field $\mib{u}(\mib{r},\tau)$ is obtained by averaging over a velocity of all the particles in a coarse grained cell at $\mib{r}$;
\begin{equation}
\mib{u}(\mib{r},\tau) \equiv
 \frac{1}{N_{\mib{r}}}
  \sum_{i}^{\mathrm{cell \; at \; } \mib{r}}\mib{v}_{i},
\end{equation}
where $N_{\mib{r}}$ is the number of the particles in the cell at $\mib{r}$. In this figure, the volume of the cell is $(1/32)^{3}$ of the whole system. In order to see the 3-d flowing structure, the velocity field is shown only in three thin layers in the system. The field is rather complex, but there can be seen some organized vortex like structure of the hydrodynamic scale.

In order to quantify the structure in the velocity field, we measure the velocity correlation function defined as
\begin{equation}
G_{\alpha\beta}(\mib{r}_{1},\mib{r}_{2}; \tau) \equiv
\langle u_{\alpha}(\mib{r}_{1},\tau)u_{\beta}(\mib{r}_{2},\tau)\rangle.
\end{equation}

In the case of the isotropic uniform system, this can be decomposed into
two parts as
\begin{equation}
G_{\alpha\beta}(\mib{r}; \tau)=
\hat{r}_{\alpha} \hat{r}_{\beta} G_{\parallel}(r,\tau)
+(\delta_{\alpha\beta} - \hat{r}_{\alpha}\hat{r}_{\beta})G_{\perp}(r, \tau),
\end{equation}
where
\begin{equation}
\mib{r} = \mib{r}_{1} - \mib{r}_{2},
\quad
\hat{\mib{r}} = \frac{\mib{r}}{r},
\end{equation}
and the longitudinal part $G_{\parallel}$ and
the transverse part $G_{\perp}$
may be calculated by
\begin{align}
G_{\parallel}(r,\tau) &=
  \langle(\mib{u}(\mib{r},\tau)\cdot\hat{\mib{r}})
  (\mib{u}(0,\tau)\cdot\hat{\mib{r}})\rangle ,      \\
G_\perp(r,\tau) &=
  \frac{1}{d-1}
\langle(\mib{u}(\mib{r},\tau)\cdot\mib{u}(0,\tau))
      -G_{\parallel}(r,\tau)\rangle .
\end{align}

These correlations have been analyzed based on the hydrodynamic equation with the Langevin force.\cite{doc1:bmesoscopic}\cite{doc1:bcahnhilliard} The actual expressions are complicated and given in Appendix\ref{doc1:as_analytical} for the 3-d case using the same notation with ref\cite{doc1:bcahnhilliard}.

\begin{figure}[htb]
\begin{center}
\scalebox{0.45}{\resizebox{\textwidth}{!}{\includegraphics{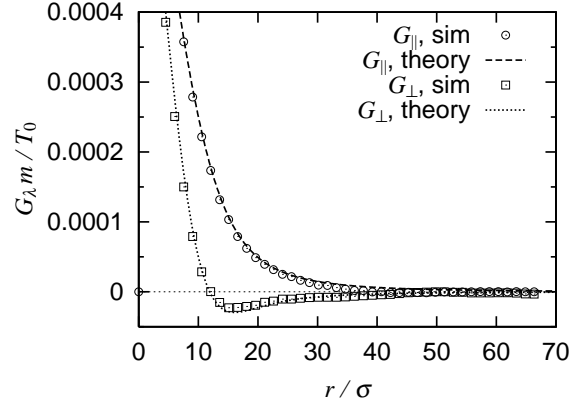}}}
\caption{The velocity correlation $G_{\parallel}$
and  $G_{\perp}$
at $\tau = 50$
for $n = 0.40,\,  e = 0.80$
and $N = 1\times 10^{6}$
just before the crossover time $\tau_{c} = 58$.
The theoretical curves are calculated for the incompressible case.}
\label{doc1:f_theory_int_tau50}

\end{center}
\end{figure}

\begin{figure}[htb]
\begin{center}
\scalebox{0.45}{\resizebox{\textwidth}{!}{\includegraphics{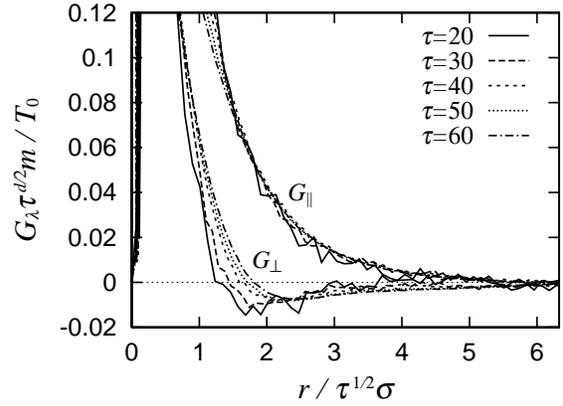}}}
\caption{Scaling form of the
velocity correlation $G_{\parallel}$
(upper curves with monotonic decay) and $G_{\perp}$
(lower curves with negative minimums)
for $n = 0.40,\,  e = 0.80$
and $N = 1\times 10^{6}$.}
\label{doc1:f_velall_scaled-large}

\end{center}
\end{figure}

Figure \ref{doc1:f_theory_int_tau50} shows $G_{\parallel}$
and $G_{\perp}$ with the theoretical curves
for the incompressible case.
The agreement is quite good.  The negative
correlation in $G_{\perp}$ is a sign of
the vortex structure.
We also evaluated the theoretical expressions of $G_{\parallel}$
and $G_{\perp}$ for the compressible
case, but the difference between the compressible and incompressible
cases are small for these parameters.

The correlations $G_{\parallel}$ and $G_{\perp}$ for several $\tau$'s are given in Fig. \ref{doc1:f_velall_scaled-large} using the scaling $G \tau^{d/2} m / T_{0}$ v.s. $r / \sigma \sqrt{\tau}$. All the plots for different $\tau$'s collapse on a single curve for $20 \leq \tau \leq 60$, which means that the correlations scale as
\begin{equation}
G(r,\tau) = \frac{T_{0}}{m\tau^{d/2}}\thinspace
 f\left(\frac{r}{\sigma \sqrt{\tau}}\right) ,
\end{equation}
very well. We did not observe large density fluctuation until $\tau \sim 70$ ($\tau_{c} = 58$).

\section{Correction for the energy decay rate}
\label{doc1:s_energy_decay_rate}
\suppressfloats[t]

As we have pointed out in \S\ref{doc1:s_hcs}, the energy decay rate in HCS slightly deviates from the one given by Haff's law. We shall examine this deviation in detail in this section.

Haff's decay rate $\gamma_{0}$ is obtained on the random collision assumption with the Gaussian distribution of the particle velocity. We will first examine the effect of the non-Gaussian velocity distribution and then the velocity correlation.

\subsection{Effect of non-Gaussian velocity distribution}

The time dependent decay rate defined as
\begin{equation}
2\gamma(\tau) = - \frac{d}{d\tau} \ln T(\tau) ,
\end{equation}
and the collision rate $\omega(\tau)$ have been studied in the case of non-Gaussian velocity distribution using the Enskog-Boltzmann theory to give \cite{doc1:bvelocity_dist_hgf}
\begin{equation}
\label{doc1:m_gamma_correction}
\frac{\gamma(\tau)}{\gamma_{0}}
 = \left[1 + \frac{3}{16}a_{2}(\tau)\right]
   \frac{\omega_{0}(\tau)}{\omega(\tau)},
\end{equation}
\begin{equation}
\frac{\omega(\tau)}{\omega_{0}(\tau)} = 1 - \frac{1}{16}a_{2}(\tau),
\end{equation}
up to the first order of $a_{2}$. Here, $a_{2}$ is given in Appendix A and $\omega_{0}(\tau)$ is the collision frequency in the Enskog-Boltzmann approximation in the equilibrium with the average velocity $v_{0}(\tau)$,
\begin{equation}
\omega_{0}(\tau)
 = \frac{\Omega_{d}\chi(n) n \sigma^{d-1}}{\sqrt{2\pi}}v_{0}(\tau),
\end{equation}
where $\Omega_{d}$ is the surface area of the $d$ dimensional unit sphere and $\chi(n)$ is the pair correlation at contact in the equilibrium with the density $n$.

\begin{figure}[htb]
\begin{center}
\scalebox{0.45}{\resizebox{\textwidth}{!}{\includegraphics{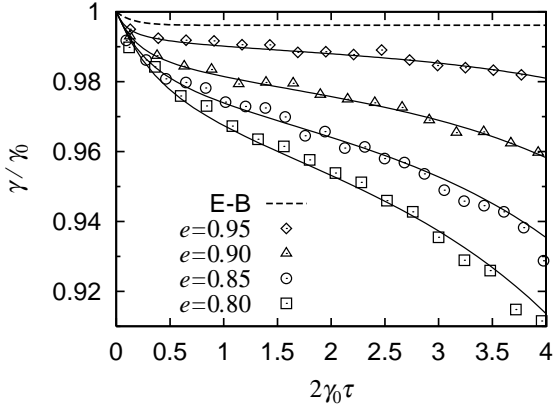}}}
\caption{Comparison of the energy decay rate $\gamma(\tau)$
of the simulation results (points) with 
theoretical estimation 
for various values of the restitution coefficient $e$
 with $n = 0.40$
and $N = 0.6\times 10^{6}$.
The dashed line represents
the Enskog-Boltzmann result eq.(\ref{doc1:m_gamma_correction})
with the analytical solution of $a_{2}$
(eq.(\ref{doc1:m_a2_analytical})) for $e = 0.865$.
Solid lines show the
phenomenological approximation
of eq.(\ref{doc1:m_energy_decay_app})
with fitting parameter $b = 0.30$.}
\label{doc1:f_gamma_int}

\end{center}
\end{figure}

Numerically obtained decay rates $\gamma(\tau)$ in the early stage of HCS are plotted in Fig. \ref{doc1:f_gamma_int} for a several values of restitution coefficient $e$. The dashed line in Fig. \ref{doc1:f_gamma_int} is the Enskog-Boltzmann prediction (\ref{doc1:m_gamma_correction}) for $e = 0.865$, for which we expect that the deviation of $\gamma$ from $\gamma_{0}$ should be largest in the range of $0.80 \leq e < 1$ because the stationary value of $|a_{2}|$ reaches the maximum at that value.

There are two points that we can see in Fig. \ref{doc1:f_gamma_int}; (i) actual deviations of $\gamma$ from $\gamma_{0}$ are much larger than this largest possible deviation within the Enskog-Boltzmann theory, and (ii) the deviation increases monotonically upon increasing dissipation while the Enskog-Boltzmann theory expects the maximum deviation around $e = 0.865$.

\begin{figure}[htb]
\begin{center}
\scalebox{0.45}{\resizebox{\textwidth}{!}{\includegraphics{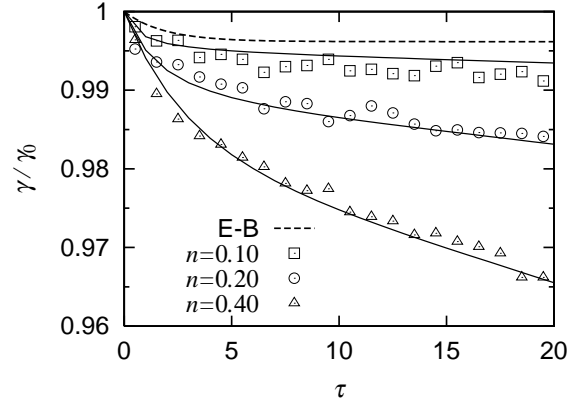}}}
\caption{Comparison of the energy decay rate $\gamma(\tau)$
of the simulation result (points) with theoretical estimates
for various values of the density $n$
with $e = 0.85$
and $N = 0.6\times 10^{6}$.
Density dependence is shown.
The dashed line represents
the Enskog-Boltzmann result eq.(\ref{doc1:m_gamma_correction})
with the analytical solution of $a_{2}$
(eq.(\ref{doc1:m_a2_analytical})).
Solid lines show the
phenomenological approximation
of eq.(\ref{doc1:m_energy_decay_app})
with fitting parameter $b = 0.30$.}
\label{doc1:f_n-gamma_int}

\end{center}
\end{figure}

Density dependence of $\gamma(\tau)$ is shown in Fig. \ref{doc1:f_n-gamma_int} for $ e = 0.85$. Although the stationary value of $a_{2}$ does not depend on the density $n$ in the theory based on the Enskog-Boltzmann equation, the simulation result of $\gamma(\tau)$ does, as well as the simulation results of $a_{2}$. The absolute value of $a_{2}$, however, becomes smaller than that of the Enskog-Boltzmann theory for larger density, while the deviation of the decay rate $\gamma$ from $\gamma_{0}$ gets larger, as $n$ increases.

These points suggest that the deviation in $\gamma$ cannot be explained by the non-Gaussian distribution, and should be mainly caused by the velocity correlation even in the early stage of HCS.

\subsection{Effect of velocity correlation}

Now, we will try to include the effect of velocity correlation. The velocity correlation results in the local average flow of particles and should lead to the following two effects that slow down the energy decay: (i) reduction of collisional dissipation and (ii) viscous heating due to shearing of the average flow. We will examine only the first one since the viscous heating is the secondary effect and should be negligible in the early stage of HCS that we are looking at.

In the presence of the local average flow $\mib{u}(\mib{r},\tau)$, the granular temperature $T_{\mathrm{L}}(\mib{r},\tau)$ should be redefined as
\begin{equation}
\label{doc1:mlocal_temperature}
T_{\mathrm{L}}(\mib{r},\tau) \equiv 
 \frac{1}{N_{\mib{r}}}
  \sum_{i}^{\mathrm{cell \; at \; }\mib{r}}
  \frac{m}{d}(\mib{v}_{i} - \mib{u})^{2},
\end{equation}
where the summation is taken over the particles in a cell at $\mib{r}$. We have to use this definition instead of the granular temperature $T$ given in eq.(\ref{doc1:maverage_temperature}), which actually corresponds to the kinetic energy.

Our physical assumption is that the collisional dissipation should be determined by the granular temperature $T_{\mathrm{L}}$ defined in eq.(\ref{doc1:mlocal_temperature}) because the average flow cannot be relaxed through local collisions. Then, Haff's law for the energy decay is extended by modifying eq.(\ref{doc1:m_temp_time}) as
\begin{equation}
\label{doc1:m_haff_modified}
\frac{dE}{d\tau}(\tau)
  = -2\gamma_{0}\left(\frac{d}{2}\langle T_{\mathrm{L}}(\tau)\rangle\right),
\end{equation}
where $E(\tau) \equiv \langle m v^{2}/2 \rangle = (d/2)T(\tau)$ is the average energy per particle.

The energy may be expressed in terms of the velocity correlation and the granular temperature as in the following. In the presence of the average flow, the kinetic energy per particle $E(\tau)$ can be decomposed into two parts as \cite{doc1:bextension_haff}
\begin{equation}
\begin{split}
E(\tau) = \frac{1}{N}\int d\mib{r}
&\left\{
 \frac{1}{2}\langle m n(\mib{r},\tau)u^{2}(\mib{r},\tau)\rangle \right. \\
&+ \left.
 \frac{d}{2} \langle n(\mib{r},\tau)T_{\mathrm{L}}(\mib{r},\tau)\rangle 
\right\},
\end{split}
\end{equation}
where $n(\mib{r},\tau)$ is the local particle density.

We assume, at this point, that $n$ and $\langle T_{\mathrm{L}} \rangle$ are uniform because we are considering early stage of HCS, then we have
\begin{equation}
\label{doc1:m_energy_tl_corr}
E(\tau) \approx \frac{1}{V}\int d\mib{r} \frac{m}{2}
\langle u^{2}(\mib{r},\tau)\rangle
+\frac{d}{2} \langle T_{\mathrm{L}}(\tau) \rangle.
\end{equation}
The first term of RHS can be expressed in terms of the velocity correlation in the hydrodynamic limit as
\begin{equation}
\begin{split}
\frac{1}{V}\int d\mib{r} \frac{1}{2}\langle u^{2}(\mib{r},\tau)\rangle 
 &= 
\frac{1}{2}\lim_{r_{12}\rightarrow 0}
  \mathrm{Tr}\left[ G(r_{12},\tau)\right], \\
 &\equiv
\frac{1}{2} \mathrm{Tr}\left[ G^{+}(0,\tau)\right] .
\end{split}
\end{equation}
The correlation function has been obtained within the linear approximation\cite{doc1:bcahnhilliard} as
\begin{equation}
\begin{split}
 \mathrm{Tr}\left[ G^{+}(0,\tau)\right]
  &= (d-1)\frac{T_{\mathrm{Haff}}(\tau)}{mn} \\
  &\times
  \int \frac{d\mib{k}}{(2\pi)^{d}}
    \frac{e^{2\gamma_{0}\tau(1 - \xi_{\perp}^{2}k^{2})} - 1}
         {1 - \xi_{\perp}^{2} k^{2}},
\end{split}
\end{equation}
for the incompressible case. The short wave length cutoff $k_{\mathrm{M}}$ is needed for the integral to give a finite result; the integral diverges logarithmically for 2-d and linearly for 3-d as $k_{\mathrm{M}} \rightarrow \infty$. We set $k_{\mathrm{M}}$ as
\begin{equation}
k_{\mathrm{M}} = 2\pi n^{1/d} \times b,
\end{equation}
where we take $b$ as a fitting parameter independent of the density and the restitution coefficient. Then we have
\begin{equation}
\label{doc1:m_correlation_integral_gamma}
\begin{split}
 \Tr\left[ G^{+}(0,\tau)\right]
  &=
  (d-1)\frac{T_{\mathrm{Haff}}(\tau)\Omega_{d}}{mn(2\pi\xi_{\perp})^{d}} \\
  &\times
    \int_{0}^{2\gamma_{0}\tau}ds \frac{e^{s}}{2s^{d/2}}
      \gamma(d/2, s\xi_{\perp}^{2} k_{\mathrm{M}}^{2}),
\end{split}
\end{equation}
where $\gamma(a,x)$ is the incomplete gamma function defined in Appendix \ref{doc1:as_analytical}. For large $\tau$, this expression gives $E \sim \tau^{-d/2}$, as has been studied in the ref\cite{doc1:bextension_haff}.

In the ref\cite{doc1:bextension_haff}, $\langle T_{\mathrm{L}} \rangle$ is assumed to obey Haff's law, but here we do not use this assumption; eliminating $\langle T_{\mathrm{L}} \rangle$ from eq.(\ref{doc1:m_haff_modified}) using eq.(\ref{doc1:m_energy_tl_corr}), a differential equation for $E(\tau)$ is obtained as
\begin{equation}
\label{doc1:m_energy_decay_app}
\frac{dE}{d\tau}(\tau)
  = -2\gamma_{0}\left(
    E(\tau) - \frac{m}{2}\mathrm{Tr}\left[G^{+}(0,\tau)\right]\right).
\end{equation}

We solve eq.(\ref{doc1:m_energy_decay_app}) numerically with
numerically evaluated $\Tr\left[G^{+}(0,\tau)\right]$ from
eq.(\ref{doc1:m_correlation_integral_gamma}).

Comparison of the energy decay rate $\gamma(\tau)$ with 
the simulation results is shown in
Figs. \ref{doc1:f_gamma_int} and
\ref{doc1:f_n-gamma_int}.
The value of the fitting parameter $b$ is chosen to be 
$b = 0.30$ to give the best fit for the energy $E(\tau)$
 before $\tau_{c}$ at $e = 0.80$ and $n = 0.40$.
Both the restitution coefficient and the density
dependence are well described in this approximation with 
the same value of the single fitting
parameter $b$.

\begin{figure}[htb]
\begin{center}
\scalebox{0.45}{\resizebox{\textwidth}{!}{\includegraphics{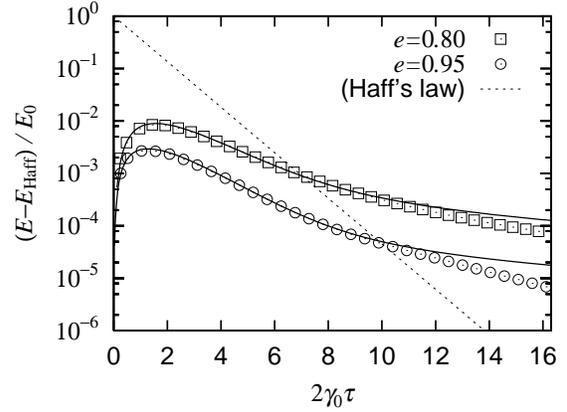}}}
\caption{The deviation of the energy $E(\tau)$
from Haff's law.
The simulation results (points) are compared 
with the phenomenological approximation of 
eq.(\ref{doc1:m_energy_decay_app}) (solid lines)
for $n = 0.40$
and $N = 0.6\times 10^{6}$.
Fitting parameter is $b = 0.30$.
The dashed line is Haff's law $\exp(-2\gamma_{0}\tau)$
(the crossing points of the simulation results and the dashed line give
 $\tau_{c}$'s in our definition in eq.(\ref{doc1:m_def_tauc})).}
\label{doc1:f_dif-int_energy}

\end{center}
\end{figure}

In Fig. \ref{doc1:f_dif-int_energy},
the deviation of the energy from Haff's law
$E(\tau) - E_{\mathrm{Haff}}(\tau)$
is plotted and the simulation results are compared with the results from
eq.(\ref{doc1:m_energy_decay_app}).
The agreement is quite good and this approximation seems to be
valid until little after $\tau_{c}$.
But in the clustering region, the simulation results have stronger energy
decay than our estimates.
These results show that the deviation of the energy from Haff's law
at the initial stage is mainly 
due to the velocity correlation, while the effect
of the viscous heating and density cluster might have to be
considered at the clustering regime.

\section{Summary and Discussions}
\label{doc1:s_summery}
\suppressfloats[t]

Unlike in the equilibrium state, the velocity correlation develops in the non-equilibrium state of the inelastic system, and it seems to play an important role in the time development of the system. We carefully looked into the early stage of the free cooling in the 3-d inelastic hard sphere system in the low dissipation range $e \geq 0.8$, and found (i) the initial energy decays exponentially, but the decay rate is slightly smaller than that expected in the case of the random collision, (ii) the velocity distribution deviates from the Gaussian but only slightly, (iii) the form of the velocity distribution does not stay stationary even in HCS contrary to the expectation by the Enskog-Boltzmann equation, (iv) the velocity field develops global structure and the velocity correlations fit to the theoretical results based on the linear analysis of the hydrodynamic equation with the Langevin force, (v) the deviation of the energy decay rate can be described in terms of the velocity correlation.

Among these results, (i), (iii), and (v) are the evidences that the velocity correlation plays an important role even in the early stage of HCS. The velocity correlation in this stage has also been studied numerically using the ring approximation of the kinetic theory\cite{doc1:bprecollisional}, where it has been shown that the velocity correlation arises within a small number of collisions.

The effects of the non-Gaussian velocity distribution and the velocity correlation on the energy decay have been also studied in the kinetic theory by P{\"o}schel \textit{et al}\cite{doc1:bviolation_molecular_chaos}. They studied the energy decay in the real time, and derived the expression for the deviation from Haff's law by the non-Gaussian distribution and the velocity correlation. By comparing it with the 2-d simulation, they concluded that a major part of the deviation should come from the velocity correlation, but they did not make theoretical estimate for the size of the velocity correlation.

The effect of the velocity correlation should depend on the particle density; the lower the density is, the smaller the correlation effect becomes, because the mean free path gets long much faster than the mean distance between particles upon decreasing $n$. However, this does not necessarily mean that the correlation effects disappear in HCS for small enough density; for large enough system, it is believed that the clustering instability takes places eventually for any density and dissipation. This should be caused by the shearing flow, and the shearing is a manifestation of the velocity correlation. Therefore, it is plausible that, even for a small density system, the correlation effect takes place from the early stage of HCS in comparison with the total period of HCS, which becomes long for small $n$.

In the present work, the simulations are done on 3-d systems and the analysis are based on the 3-d expressions, but the above findings are not particular to 3-d, and the physical processes do not seem to be very different from 2-d system, especially in the early stage where linear theories hold. As for the late stage cluster growth, some differences between 2-d and 3-d system have been reported from MD simulations \cite{doc1:bcluster_growth}.

Then shouldn't there be any differences between 2-d and 3-d cases before the clustering?

In the 2-d system, the velocity correlation developed in the system has been visualized as a vortex structure in the intermediate stage of HCS. The appearance of the structure in 2-d reminds us of the vortex structure in the free decaying turbulence,\cite{doc1:b2d_granular_turbulence} where the dual cascade of energy and enstrophy is a leading process in the inertial range. In the 3-d system studied here, we have not succeeded in visualizing the clear vortex structure although there seems to be some global structure. If the dynamics of vortices controls the system behavior in the later stage of HCS, or the shearing state, there may be an important differences between the two cases; in 2-d, the vorticity $\mib{\omega}
 = \mib{\nabla}\times\mib{u}$ has only one component and the coarsening of the scaler field is a major process, but in 3-d, the vortices are lines in three dimensional structure and their motion such as stretching and bending accompanied by reconnection of vortex lines could cause more complex dynamics that keeps certain fine structures in the late stage; this is not taken in the linear theory. Such dynamics should have some effects in the early stage of clustering because the clustering is known to be triggered by the viscous heating in the vortex structure.

\appendix

\section{Second order approximation for Laguerre expansion coefficients  in higher dimensions}
\label{doc1:as_laguerre_coeff}
\suppressfloats[t]

In this appendix, we present the expression of $a_{2}$ within the Enskog-Boltzmann equation for the second order \cite{doc1:bgaussian} for $d$ dimensions. $a_{2}$ follows the equation,
\begin{equation}
\frac{d a_{2}}{d\tau}(\tau) = A + a_{2}(\tau)B,
\end{equation}
with
\begin{equation}
A=\frac{(e^{2} - 1)(2e^{2} - 1)}{d(d + 2)},
\end{equation}
\begin{equation}
B= -\frac{(1+e)(9+24d+8ed-41e+30(1-e)e^{2})}{16d(d + 2)}.
\end{equation}
Its solution is
\begin{equation}
\label{doc1:m_a2_analytical}
a_{2}(\tau) = -\frac{A}{B}(1 - \exp(B\tau)),
\end{equation}
and stationary value is
\begin{equation}
a_{2}^{*} = -\frac{A}{B}.
\end{equation}

\section{Analytical form of velocity correlation function}
\label{doc1:as_analytical}
\suppressfloats[t]

In the ref.\cite{doc1:bcahnhilliard}, the velocity correlation functions have been obtained from the hydrodynamic equations with the linear approximation for both incompressible and compressible flows in the form of the Fourier integrals, but the explicit expressions for $d$ dimensions are given only for the incompressible case. The expressions for $d$ dimensions used in the text are given by
\begin{equation}
\begin{split}
G_{\lambda}^{+}
 &= \frac{T_{\mathrm{Haff}}}{mn}
  \int_{0}^{2\gamma_{0}\tau}ds'
  e^{s'} \\
 \times
 &\left[
  \sigma_{\lambda}\frac{(d-1)^{\delta_{\parallel \lambda}}}{2\pi^{d/2}r^{d}}
  \left(\gamma(\frac{d}{2}, \frac{x_{\perp}^{2}}{4s'})\right) \right. \\
 &\quad + \left.
  \delta_{\perp \lambda}\frac{1}{\xi_{\lambda}^{d}}e^{-x_{\lambda}^2/4s'}
   \frac{1}{(4\pi s')^{d/2}}
 \right],
\end{split}
\end{equation}
for the incompressible fluid and
\begin{equation}
\begin{split}
G_{\lambda \mathrm{c}}^{+} 
 &= \frac{T_{\mathrm{Haff}}}{mn}
  \int_{0}^{2\gamma_{0}\tau}ds'
  e^{s'} \\
 \times
 &\left[
  \sigma_{\lambda}\frac{(d-1)^{\delta_{\parallel \lambda}}}{2\pi^{d/2}r^{d}}
  \left(\gamma(\frac{d}{2}, \frac{x_{\perp}^{2}}{4s'})
      -\gamma(\frac{d}{2}, \frac{x_{\parallel}^{2}}{4s'})\right)\right.\\
 &\quad + \left. \frac{1}{\xi_{\lambda}^{d}}e^{-x_{\lambda}^2/4s'}
\frac{1}{(4\pi s')^{d/2}}
 \right],
\end{split}
\end{equation}
for the compressible fluid, where $\lambda = \{\parallel, \perp \},
\thinspace \delta_{\lambda \lambda'}$ is Kronecker's delta, $\sigma = \delta_{\parallel \lambda}
 - \delta_{\perp \lambda}$, $T_{\mathrm{Haff}}$ is the temperature given by Haff's law (eq.(\ref{doc1:m_haff_tau})),
\begin{equation}
\gamma(a, x) = \int_{0}^{x}dt \exp(-t)t^{a-1},
\end{equation}
is the incomplete gamma function,
\begin{equation}
x_{\lambda} = r/\xi_{\lambda} \, ,
\end{equation}
and $\xi_{\lambda}$'s are the correlation lengths of the elastic hard sphere system, given in the ref\cite{doc1:bcahnhilliard}.

\bibliographystyle{plain}
\bibliography{ronbun_master}
\nocite{doc1:bhaff}
\nocite{doc1:bkinetic_theory}
\nocite{doc1:bmesoscopic}
\nocite{doc1:bcahnhilliard}
\nocite{doc1:bsimple_efficient_algorithm}
\nocite{doc1:bextension_haff}
\nocite{doc1:bgaussian}
\nocite{doc1:blaguerre2d}
\nocite{doc1:bvelocity_dist_hgf}
\nocite{doc1:borigin_of_density_cluster}
\nocite{doc1:btwo_dim_regime}
\nocite{doc1:binelastic_collapse}
\nocite{doc1:bviolation_molecular_chaos}
\nocite{doc1:bprecollisional}
\nocite{doc1:bcluster_growth}
\nocite{doc1:b2d_granular_turbulence}

\end{document}